\documentstyle[preprint,aps]{revtex}
\tightenlines

\begin{document}

\draft
\preprint{INJE-TP-00-2, hep-th/0001211}

\title{ Schwarzschild black hole in the
dilatonic domain wall}

\author{ Y.S. Myung and H.W. Lee}
\address{Department of Physics, Inje University, Kimhae 621-749, Korea}

\maketitle

\begin{abstract}
In  the dilatonic domain wall model, we study the Schwarzschild black hole as a
solution to the Kaluza-Klein (KK) zero
mode effective action which is equivalent to the Brans-Dicke (BD) model with
a potential.
This  can describe   the large Randall-Sundrum (RS) black hole
whose horizon
is to be  the intersection  of  the black
cigar with the brane.
 The black cigar located far from the AdS$_5$-horizon  is known  to be stable, but
any  explicit calculation for stability of the RS black hole at $z=0$
is not yet performed.
Here its stability  is investigated against
the $z$-independent perturbations composed of odd, even parities of graviton ($h_{\mu\nu}$) and
 BD scalar($h_{44} = 2\varphi $).
It seems that the RS black hole is classically unstable
because it has a potential instability at wavelength with $\lambda > 1/(2k)$.
However,  this is not allowed inside an AdS$_5$-box of the size
with $1/(2k)$. Thus the RS black hole becomes stable.
The  RS black hole can be considered as a stable
remnant
at $z=0 $ of the black cigar.
\end{abstract}
\vfill
%Compiled at \today : \number \time.

\newpage
\section{Introduction}
\label{introduction}
Recently there has been much interest in the Randall-Sundrum
brane world\cite{Ran99PRL3370,Ran99PRL4690,Ark9907209}.
A key idea of this model is that our universe may be a brane embedded in
the higher dimensional space. A concrete model is a single 3-brane embedded
in the five-dimensional anti-de Sitter space (${\rm AdS}_5$), which acts like a box of the size
$\sim 1/k$.  Randall and
Sundrum have shown that a longitudinal part ($h_{\mu\nu}$) of the
metric fluctuations satisfies the Schr\"odinger-like equation with an
attractive delta-function. As a result, the massless KK  modes which
describe the localized gravity on the brane were found. Furthermore,
the massive KK modes lead to  corrections to the Newtonian potential
like as $V(r) = G_N \frac{m_1m_2}{r}(1+\frac{1}{k^2r^2})$.

However, we would like to point out that this has been done in the 4D Minkowski space
with the RS gauge\footnote{In fact, this gauge for $h_{MN}$ is
composed of Gaussian-Normal (GN) gauge ($h_{44} = h_{4\mu}=0$)
and 4D transverse, tracefree (TTF) gauge($\partial^\mu h_{\mu\nu} =0$,
$h^\mu_{~\mu} = 0$).}.
It seems that this gauge is so restrictive. In order to
have  well-defined theory on the brane, one has to include
non-zero transverse parts of $h_{4\mu}$ and $h_{44}$ at the beginning.
Ivanov and Volovich discussed along this direction by choosing
the 5D de Donder gauge\footnote{This corresponds to the 5D TTF gauge
($\partial_M h^{MN} =0, h^M_{~M} = 0$).}\cite{Iva9912242,Myu0001003}.
Also authors in Ref.\cite{Myu0001107} studied
  the  propagation of the metric including with non-zero transverse parts.
  It turned out that there are no massless scalar and
 vector propagations on the RS background.
This implies that the RS gauge is a good choice for describing the brane world.
Furthermore, it is shown that the RS Minkowski spacetime  is
stable only under the RS gauge.

If the  Minkowski metric on the brane is replaced by any
4D Ricci-flat one, for example the Schwarzschild metric, the 5D
metric still satisfies the Einstein equation with a negative
cosmological constant.
In this process we obtain  a black string  solution in  an AdS$_5$
\footnote{Precisely, this is not a  5D black hole solution to the RS model.
At present, nobody knows that.}\cite{Cha9909205}.
This is given by ${\bar G}_{MN}^{\rm BS} = H^{-2}(z)[\bar g_{\mu\nu}^S,1]
$ with the Schwarzschild metric  $\bar g^S_{\mu\nu}={\rm diag}[
-(1-2M/r), (1-2M/r)^{-1}, r^2, r^2 \sin^2\theta]$.
It is shown that this  black string is unstable near the AdS$_5$
horizon of  $z=\infty$ but it is stable far from the horizon.
This is a result of  the Gregory-Laflamme instability\cite{Gre93PRL2839}, which
states that the black string has a tendency to fragment near
the AdS$_5$-horizon\cite{Gre0004101}.
A stable object left behind would resemble a black cigar, although we do not know
its explicit metric.
Hence we have the RS black hole-picture on the brane located at
$z=0$.
The horizon of  black hole on the brane
will be determined  from the intersection of the black cigar with the brane.
Here we assume that this  is  large  such as $r_{EH}=2M>
1/k$.
In this case an observer on the brane perceives exactly
the  Schwarzschild solution, without any correction arising
from the extra dimension\cite{Emparan}. In other words, any massive KK mode
are not excited in  this circumstances . Then the zero modes of the bulk graviton
can describe this situation very well. Hence the zero mode approach
becomes a powerful technique in the study of  black holes on the brane.

On the other hand, Youm showed that the RS solution can be found
in the dilatonic domain wall\cite{You0001018,You0001166}.
In order to study physics on the brane, we need its effective action.
It is known that zero mode effective action takes a form of   the Brans-Dicke model with
a potential.

In this paper, we study  stability of the large RS black hole
with the $z$-independent perturbations such as
$h_{\mu\nu}(x)$, $h_{44}(x) $.
We find that these
are massless  graviton and scalar
 modes propagating in the RS  black hole background.
Here we do not require  the  4D TTF gauge
 which is useful for the RS Minkowski space.
Instead we choose the Regge-Wheeler (RW) gauge for our study of
 spherically symmetric background.

\section{Randall-Sundrum solution and Brans-Dicke type model}
\label{sec-randall}
We start with the 5D bulk action and
the 4D domain wall action  as\cite{You0001166}
\begin{equation}
 S = S_{\rm bulk} + S_{\rm DW}
\label{action}
\end{equation}
with
\begin{eqnarray}
S_{\rm bulk} &=& {1 \over { 2 \kappa_5^2}} \int d^5x \sqrt{-G}
  \left [ R_5 - {4 \over 3} \partial_M D \partial^M D
  - e^{-2 a D} \Lambda \right ],
\label{actionbulk} \\
S_{DW} &=& - \sigma_{\rm DW} \int d^4 x \sqrt{-\gamma} e^{-a D} ,
\label{actiondw}
\end{eqnarray}
where $\sigma_{\rm DW}$ is the tension of the domain wall and $\gamma$
is the determinant of the  induced metric
$\gamma_{\mu\nu} = \partial_\mu X^M \partial_\nu X^N G_{MN}$ for  the
domain wall.
Here $M,N=0,1,2,3,4(x^4=z)$ and $\mu, \nu = 0,1,2,3(x^\mu = x)$.
``$D$'' denotes the dilaton.
We are interested in the second RS solution with\footnote{With this ansatz,
the dilatonic action Eq.(\ref{action}) reduces to  the second RS
model exactly\cite{Ran99PRL4690}}

\begin{equation}
{\bar G}_{MN} = H^{-2}(z) \eta_{MN}, ~~ \bar D=0,
~~ \Lambda = -12 k^2, ~~ \sigma_{\rm DW} = 6 k / \kappa_5^2 , ~~ a=0
\label{rssolution}
\end{equation}
with $H= k |z|+1$ and $\eta_{MN} = {\rm diag}[-++++]$.
Here overbar($^-$) means the background value.
In this paper we follow the MTW conventions\cite{Mis73G}.

In order to obtain a 4D effective action, we introduce the
metric ${\hat G}_{MN}$ in $G_{MN} = H^{-2}(z) {\hat G}_{MN}$ which
satisfies $\partial_z {\hat G}_{MN} = 0$.
Explicitly, the line element is given by
\begin{eqnarray}
dS_5^2 &=& G_{MN} dx^M dx^N = H^{-2} {\hat G}_{MN} dx^M dx^N
\nonumber \\
   &=& H^{-2} \left [ g_{\mu\nu}(x) dx^\mu dx^\nu
    + \Phi^2(x) dz^2 \right ].
\label{5metric}
\end{eqnarray}
Off-diagonal elements are not allowed because if they exist, these violate the
$Z_2$-symmetry argument\footnote{One may introduce  off-diagonal term
of $2A_\mu(x)dx^\mu dz$. But this is not invariant under the reflection of $z \to -z $
\cite{Ran99PRL3370}. Thus  we  must  have  $A_\mu(x)=0$.}.
At this stage we wish to remind the reader that  $\Phi(x)$ has nothing
to do with the radion which is necessary for stabilizing the distance between two branes
 in the first RS model\cite{Ran99PRL3370}.
This is because here we consider  the second RS model\cite{Ran99PRL4690}.
Substituting (\ref{5metric}) with $D=0$ into (\ref{action}) and
integrating it over $z$ lead to the Brans-Dicke type model with a
potential\cite{You0001166}
\begin{equation}
S_{\rm RS} = {1 \over {2 \kappa_4^2}} \int d^4 x \sqrt{-g} \left [
\Phi R + 6 k^2 \left ( \Phi + { 1\over \Phi} -2 \right ) \right ]
\label{actionrs}
\end{equation}
with $\kappa_4^2 = k \kappa_5^2$.
This is our key action which is suitable for the study of physics defined on the brane.
Also it can be derived from the RS model directly.
It is emphasized that the first term($\Phi R$) is
 the BD term with $\omega=0$\cite{Bra61PR925}.
The BD model with $\omega=0$ corresponds to the massless Kaluza-Klein
model with $g_{\mu\nu}$, $g_{44}(\sim \Phi^2)$, and
$g_{\mu4}(\sim A_\mu)=0$~\footnote{
This model is based on the action of
$S_{\rm KK} = {1 \over \kappa_5^2} \int d^5 x \sqrt{-G} R_5$ with
 a factorizable geometry of $H=1$.}.
In this sense, we wish to call $\Phi$ as the BD scalar.
The second term  arises from the fact that
the 5D spacetime is an  AdS$_5$ with a negative cosmological constant $\Lambda$ and
a domain wall located at $z=0$.
Equivalently, this means that the RS solution gives us  a
non-factorizable geometry with $H= k|z| +1$.
Hence this term accounts for  the feature
of the RS type solution and it plays the  role of an effective potential.
>From now on we wish to separate the pure BD model
(${\cal L}_{\rm BD} = \Phi R$) from the RS model
(${\cal L}_{\rm RS} = \Phi R + 6 k^2(\Phi + 1/\Phi -2 $)) for comparison.

>From (\ref{actionrs}) we derive the equations of motion
($\delta_\Phi S_{\rm RS} =0, \delta_{g_{\mu\nu}} S_{\rm RS} =0$)
\begin{eqnarray}
&& R + 6 k^2 \left ( 1 - {1 \over {\Phi^2}} \right ) =0,
\label{eq-r} \\
&& R_{\mu\nu} - {1 \over 2} g_{\mu\nu} R =
{1 \over \Phi} \left \{ \nabla_\mu \nabla_\nu \Phi
  - g_{\mu\nu} \Box \Phi + 3 k^2 \left ( \Phi + {1 \over \Phi} -2 \right )
  g_{\mu\nu} \right \} .
\label{eq-rmunu}
\end{eqnarray}
Contracting Eq.(\ref{eq-rmunu}) with $g^{\mu\nu}$ and using
Eq.(\ref{eq-r}) leads to the RS scalar equation
\begin{equation}
\Box \Phi + 2 k^2 \left ( -\Phi - {3 \over \Phi} +4 \right ) =0.
\label{eq-bd}
\end{equation}
Also from Eq.(\ref{eq-rmunu}), its contraction form, and
(\ref{eq-bd}) one finds the other Einstein equation
\begin{equation}
 R_{\mu\nu} = {1 \over \Phi} \left \{ \nabla_\mu\nabla_\nu \Phi
  +  2 k^2 \left ( -\Phi + 1  \right ) g_{\mu\nu}
   \right \}.
\label{eq-rmunu1}
\end{equation}
Here one finds a solution which satisfies all of
equations (\ref{eq-r})-(\ref{eq-rmunu1}) simultaneously as
\begin{equation}
\bar \Phi =1, ~~ \bar R =0, ~~ {\bar R}_{\mu\nu} =0.
\label{flatsolution}
\end{equation}
This means that the Ricci-flat condition of  ${\bar R}_{\mu\nu}=0$
with ${\bar \Phi}=1$ describes the 4D vacuum configuration correctly.

As an example, we choose the Minkowski spacetime
\begin{equation}
{\bar g}_{\mu\nu} = \eta_{\mu\nu}, ~~ {\bar \Phi} = 1.
\label{4dminkowski}
\end{equation}
To study the propagations on this background, we introduce the perturbation around the
background (\ref{4dminkowski}) as
\begin{equation}
g_{\mu\nu} = \eta_{\mu\nu} + h_{\mu\nu}, ~
\Phi = {\bar \Phi} + \varphi~(g_{44} = {\bar g}_{44} + h_{44} ).
\label{ptr-minkowski}
\end{equation}
Then the linearized equations to (\ref{eq-bd}) and (\ref{eq-rmunu1}) are
found to be
\begin{eqnarray}
&&\partial^2 \varphi + 4 k^2 \varphi = 0 ,
\label{ptr-phi} \\
&& \delta R_{\mu\nu}(h) - \partial_\mu \partial_\nu \varphi
+ 2 k^2 \varphi \eta_{\mu\nu} = 0
\label{eq-deltar}
\end{eqnarray}
with
\begin{equation}
\delta R_{\mu\nu} = -{1 \over 2} \left [
\partial^2 h_{\mu\nu} + \partial_\nu \partial_\mu h^\rho_{~\rho}
- \partial^\rho \partial_\mu h_{\nu\rho}
- \partial^\rho \partial_\nu h_{\mu\rho}
\right ].
\label{deltar}
\end{equation}
Under the 4D TTF gauge, one finds that
$\delta R_{\mu\nu} = -{1\over 2} \partial^2 h_{\mu\nu}$ and
$\eta^{\mu\nu} \delta R_{\mu\nu} = - {1 \over 2} \partial^2 h^\mu_{~\mu} = 0.$
Contracting Eq.(\ref{eq-deltar}) with $\eta^{\mu\nu}$ leads to
the other equation for $\varphi$
\begin{equation}
\partial^2 \varphi - 8 k^2 \varphi = 0.
\label{cont-phi}
\end{equation}
Eq.(\ref{ptr-phi}) allows a tachyonic solution because
it has a negative potential term of  $-4k^2$.
Fortunately  we resolve this problem.
We have two different equations (\ref{ptr-phi}) and (\ref{cont-phi})
for the same field of $\varphi$.
Hence we require $\varphi=0$ for  consistency.
This observation agrees  with Refs.\cite{Myu0001107,You0001166}.
Then Eq.(\ref{eq-deltar}) reduces to an
equation for the massless graviton without a matter source on the brane
\begin{equation}
\partial^2 h_{\mu\nu} = 0.
\label{eq-massless}
\end{equation}

\section{Schwarzschild black hole solutions}
\label{sec-schwarzschild}
Introducing a  spherically symmetric spacetime, one obtains
the Schwarzschild black hole
with $\bar\Phi=1$ in the domain wall approach\cite{Cha9909205} as
\begin{eqnarray}
&&{\bar g}_{\mu\nu} = {\rm diag} \left [ -e^\nu, e^{-\nu},
   r^2, r^2 \sin^2 \theta \right ],
\label{sol-gmunu} \\
&&
{\bar R}_{trtr} = 2 {M \over r^3}, ~~
{\bar R}_{t\theta t\theta} = - {{(r -2 M) M} \over r^2}, ~~
{\bar R}_{t\phi t\phi} = \sin^2\theta {\bar R}_{t\theta t\theta}, ~~
\nonumber \\
&&
{\bar R}_{r\theta r\theta} = {M \over {r -2 M}}, ~~
{\bar R}_{r\phi r\phi} = \sin^2\theta {\bar R}_{r\theta r\theta}, ~~
{\bar R}_{\theta\phi \theta\phi} = -2 Mr\sin^2\theta, ~~
\nonumber \\
&& {\bar R}_{\mu\nu} =0, ~~ \bar R =0
\nonumber
\end{eqnarray}
with
\begin{equation}
e^\nu = 1 - {2M \over r}.
\label{enu}
\end{equation}
Here $M = G_N M_4 = \kappa_4^2 M_4 / 8 \pi = G_5 k M_4$ with
$G_N$ (4D Newtonian constant) and $G_5$ (5D Newtonian constant).
$M_4$ is the mass of a large black hole with $M_4 > 1/(2k^2G_5)$.
 The BD and RS black hole solutions are permitted because the
Schwarzschild solution comes from the Ricci-flat condition.
 The BD (RS) black holes denote the Schwarzschild
black hole with ${\cal L}_{\rm BD}({\cal L}_{\rm RS})$, respectively.
The BD black is introduced for reference.
To study these black holes specifically, we introduce the perturbation
\begin{equation}
g_{\mu\nu} = {\bar g}_{\mu\nu} + h_{\mu\nu}, ~~ \Phi = \bar\Phi + \varphi.
\label{def-perturb}
\end{equation}
Then the linearized equations to Eqs. (\ref{eq-bd}) and (\ref{eq-rmunu1}) are
found as
\begin{eqnarray}
&& \bar{\lower.4ex\hbox{$\Box$}} \varphi + 4 k^2 \varphi =0,
\label{eq-phi} \\
&& \delta R_{\mu\nu}(h) - \bar\nabla_\nu \bar\nabla_\mu \varphi
   + 2 k^2 \varphi {\bar g}_{\mu\nu} =0
\label{eq-h}
\end{eqnarray}
with the Lichnerowicz operator $\delta R_{\mu\nu}(h)$\cite{Lee98PRD104006,Myu9912288}
\begin{equation}
\delta R_{\mu\nu}(h) = -{1 \over 2} \left [ {\bar{\lower.4ex\hbox{$\Box$}}} h_{\mu\nu}
    + \bar \nabla_\nu\bar\nabla_\mu h^\rho_{~\rho}
    - \bar\nabla^\rho \bar\nabla_\mu h_{\nu\rho}
    - \bar\nabla^\rho \bar\nabla_\nu h_{\mu\rho} \right ].
\label{lichnero1}
\end{equation}
We note that Eq.~(\ref{eq-h}) is not a diagonalized form to
obtain  eigenmodes. If one introduces
${\hat h}_{\rho\nu} = h_{\rho\nu} +  \varphi {\bar g}_{\rho\nu}$,
then this leads to
\begin{equation}
\delta R_{\mu\nu}(\hat h) = 0.
\label{R-eq}
\end{equation}
This is the perturbed equation of  pure 4D gravity for $\hat
h_{\mu\nu}$\footnote{Its stability analysis was performed 30 years
ago\cite{Reg57PR1403,Cha92MTBH}
.}. A  way to analyze Eqs.(\ref{eq-phi}) and (\ref{eq-h}) is  known
\cite{Kwo86PRD333,Kwo86IJMP709}.
For instance, it is possible if one uses  the RW gauge instead of
the  4D TTF gauge.
The perturbation $\hat h_{\mu\nu}$ falls into two distinct classes - odd and
even parities with $(-1)^{l+1}$ and $(-1)^l$, respectively.
$l$ denotes an angular quantum number on $S^2$:
${\bar L}^2 Y_{lm}(\theta, \phi) = -l(l+1) Y_{lm}(\theta,\phi)$.
Among ten components in the axisymmetric perturbation, one
can always choose six components  by
taking into account the general coordinate transformations
: ${x^\mu}' = x^\mu + \epsilon \xi^\mu$\cite{Gar9909005,Nic0001021}.
This is a choice of the RW gauge. And this  is obvious here because we consider
the propagation of gravitons on the brane.
In the RW gauge we assign two components ($h_0, h_1$) for odd parity
\begin{equation}
\hat h_{\mu\nu}^{\rm odd} = \left (
\begin{array}{cccc}
0 & 0 & 0 & h_0(r) \\
0 & 0 & 0 & h_1(r) \\
0 & 0 & 0 & 0 \\
h_0(r) & h_1(r) & 0 & 0
\end{array}
\right ) e^{-i\omega t} \sin\theta {{d P_l(\theta)} \over {d
\theta}}
\label{h-odd}
\end{equation}
with Legendre polynomial $P_l(\theta)$.
For even parity, we have $\hat h^{even}_{\mu\nu}= h^{even}_{\mu\nu} +
\varphi \bar g_{\mu\nu}$, where $h^{even}_{\mu\nu}$ is composed of
four components ($H_0, H_1, H_2, K$) as
\begin{equation}
h_{\mu\nu}^{\rm even} = \left (
\begin{array}{cccc}
H_0\left ( 1 - {2M\over r} \right ) & H_1 & 0 & 0 \\
H_1 & H_2\left ( 1 - {2M\over r} \right )^{-1} & 0 & 0 \\
0 & 0 & r^2 K(r) & 0 \\
0 & 0 & 0 & r^2\sin^2\theta K(r)
\end{array}
\right ) e^{-i\omega t} P_l(\theta).
\label{h-even}
\end{equation}
These two cases are never mixed and thus they provide two degrees of freedom
 which is necessary for describing a massless spin-2 particle.

\subsection{RS  scalar perturbation}
\label{sec-bdscalar}
Let us first analyze the RS scalar perturbation using Eq.(\ref{eq-phi}).
Considering
$\varphi \propto {\psi(r)\over r} Y_{lm}(\theta, \phi) e^{-i\omega t}$
and the background (\ref{sol-gmunu}), one finds the
Schr\"odinger-type equation
\begin{equation}
{{d^2 \psi} \over {d {r^*}^2}} + \left ( \omega^2 - V_\psi^{\rm RS}
        \right ) \psi =0,
\label{eq-schrodinger}
\end{equation}
where the RS scalar potential is given by
\begin{equation}
V_\psi^{\rm RS} (r) = \left ( 1 - {2M \over r} \right )
  \left \{ {{l(l+1)} \over r^2 } + {2M \over r^3 } - 4 k^2
  \right \}
\label{potential}
\end{equation}
with the tortoise coordinate
$r^* = r + 2 M \ln \left ( { r \over 2M} -1 \right )$.
>From the analysis in ref.\cite{Kwo86IJMP709}, we find
that for $\omega = i\alpha, 0 < \alpha < 2 k $,
the scalar perturbation has an exponentially growing mode of
$e^{\alpha t}$. Therefore this system may be classically unstable.
In other words, ``$-4k^2$'' in the potential induces an instability of the asymptotically
flat space of $r\to\infty$. Thus we call it as a potential
instability.
This may imply that the RS black hole solution is classically
unstable.
However, it is not true.
Any  exponentially growing mode is  not allowed for the RS
 spherically symmetric background. An important point what we remind is that
  an AdS$_5$ acts like a box of
 size\footnote{Although a conventional length scale of an AdS$_5$ is
 determined by $\int d^5x \sqrt{-G} R_5$ in Eq.(\ref{actionbulk}) as
 $\int dz H^{-3}=1/k$,
 we here choose  a size of the  AdS$_5$-box  as $1/(2k)$
 for a definite calculation. This comes from :$-\Lambda \int d^5x \sqrt{-G} \to
 \int dz H^{-5}= 1/(2k)$.} $\sim (2k)^{-1}$
 \cite{Cha9909205}. So it may  allow an unstable perturbation of
 wavelength with $\lambda \leq (2k)^{-1}$ only\cite{CCEH}. However, for this case, we cannot
 find any consistent (unstable) solution\cite{Kwo86IJMP709}. In  the unstable  case of
  $0 < \alpha=\lambda^{-1} < 2k$, one finds a condition of $\lambda > (2k)^{-1}$
  which is forbidden inside  an AdS$_5$. The instability problem
  can be  cured by considering the bulk spacetime.
 Hence we can include the  RS  scalar mode, as a physical field,
 which  propagates in the RS
 black hole background.

\subsection{Odd parity perturbation for  $\hat h_{\mu\nu}$}
\label{sec-odd}
Now we discuss the odd parity perturbation for $\hat h_{\mu\nu}$.
>From Eq.(\ref{eq-h}), we have three equations :
\begin{equation}
\delta R_{03}=\delta R_{13} = \delta R_{23} =0.
\end{equation}
Here we obtain  the Regge-Wheeler equation using
$Q\equiv { h_1 \over r} \left ( 1 - {2M\over r} \right )$
\begin{equation}
{{d^2 Q} \over { d {r^*}^2}} + \left ( \omega^2 - V_{\rm RW}(r) \right ) Q =0,
\label{eq-rw}
\end{equation}
where the Regge-Wheeler potential is given by
\begin{equation}
V_{\rm RW}(r) = \left ( 1 - {2M\over r} \right )
  \left ( {{l(l+1)} \over r^2 } - {6M \over r^3} \right ).
\label{potential-rw}
\end{equation}
Further $h_0$ is not an independent mode,  it can be  expressed in terms of $Q$ as
$h_0 = {i \over \omega} { d \over {d r^*}} (r Q)$.
We note that the  BD black hole\cite{Kwo86PRD333}
takes the same potential
as in Eq.(\ref{potential-rw}) for the odd parity perturbation.
This case has already been analyzed by Vishveshwara in ref.\cite{Reg57PR1403}
and an  allowed solution is a scattering state
\begin{eqnarray}
Q_\infty &=& e^{-i\omega r^*} + A^- e^{i\omega r^*} ~(r^* \to \infty ),
\nonumber \\
Q_{2M} &=&  B^- e^{-i\omega r^*} ~(r^* \to -\infty ).
\label{sol-infty}
\end{eqnarray}

\subsection{Even parity perturbation for $\hat h_{\mu\nu}$}
\label{sec-even}
In this case, from the remaining seven equations, we have
 the Zerilli equation in ref.\cite{Reg57PR1403}
\begin{equation}
{{d^2 \psi_{\rm Z}} \over {d {r^*}^2}} +
  \left ( \omega^2 - V_{\rm Z} (r) \right ) \psi_{\rm Z} =0,
\label{eq-zerilli}
\end{equation}
where the Zerilli potential  is given by
\begin{equation}
V_{\rm Z}(r) = \left ( 1 -{2M\over r} \right ) \left \{
\frac{2 \lambda^2(\lambda+1) r^3 + 6\lambda^2 M r^2 + 18\lambda M^2 r + 18 M^3}
{r^3(\lambda r + 3 M )^2}
\right \}
\label{potential-zerilli}
\end{equation}
with $\lambda = (l-1)(l+2)/2$.
Here $\psi_{\rm Z}(r)$ is a gauge invariant combination of
$H_0, H_1, H_2, K, \psi/r$\cite{Kwo86PRD333,Gar9909005,Nic0001021}.
At this stage we would like to comment that the BD  equation
($\delta R_{\mu\nu}(h) - \nabla_\nu\nabla_\mu \varphi =0$)
in ref.\cite{Kwo86PRD333} takes the same  equation as
in Eq.(\ref{eq-zerilli}).
Also it is easily shown that an allowed solution is a plane wave
like  Eq.(\ref{sol-infty}).
This fact can  be easily read off from the shape of potentials $V_{\rm RW}$ and
$V_{\rm Z}$.
Because these all belong to  positive potential barrier for $l \ge 2$,
there exist scattering states only.
In other words, there are no bound state solutions.
This means that  one cannot find any exponentially
growing mode  in the graviton sector, even if $k^2$-term is involved in the RS
 equation (\ref{eq-h}).

\section{Discussions}
\label{sec-discussions}
We investigate the zero mode sector of the 5D dilatonic domain wall
solution.
This sector is very useful for describing the RS black hole on the brane.
Assuming a spherically symmetric spacetime, one has a
large   black hole on the brane.
We perform the analysis of stability to see whether or not the
RS black hole truly exists. It is well-known that
the BD black hole is stable.
Here one finds an exponentially growing mode for the RS black hole
because of a negative nature of its potential.
If there is an exponentially growing mode, its black hole  is classically unstable.
>From the analysis of $\varphi$ in  the RS Minkowski
spacetime, we have $\varphi=0$ under the 4D TTF gauge.
On the other hand, we cannot have $\varphi=0$ in  the spherically
symmetric  black hole spacetime with the RW gauge.
Fortunately,  we find that this  instability is not allowed inside an AdS$_5$ whose  size
is $(2k)^{-1}$. Hence the instability problem is cured.
Actually  BD scalar as well as  gravitons  can propagate in the  black hole spacetime.
Also this implies that the large RS black hole is  stable.

For the metric perturbations, we don't worry about their
stability even for considering $\varphi \neq 0$.
Choosing the RW gauge, the graviton sector including $\varphi$ leads to the well-known
two classes of odd and even-parities.
Since this sector has always positive potential barriers
for any $l$ with $l \ge 2$\cite{Cha92MTBH},
 there are no exponentially growing modes.
Also this can be confirmed from the other side. If we introduce
a new tensor $\hat h_{\mu\nu}$, Eq.~(\ref{eq-h}) reduces to
Eq.~(\ref{R-eq}).
This is nothing but the perturbed equation for  pure 4D
gravity, which was proved to be stable thirty years ago.

In conclusion, the large black hole in the dilaton domain wall is stable.
This can represent a stable form on the brane for the RS black cigar whose metric is not
 known up to now.
Finally we comment that
the RS black hole  can be described by  massless gravitons and
a scalar mode with smaller  wavelength  than the size of  an  AdS$_5$-box.
\section*{Acknowledgement}
We thank to J.Y. Kim and G. Kang for hepful discussions.
This work was supported in part by the Brain Korea 21
Program of  Ministry of Education, Project No. D-0025 and
 KOSEF, Project No. 2000-1-11200-001-3.

\end{document}